\documentclass[twocolumn,showpacs,preprintnumbers,amsmath,amssymb]{revtex4}
\usepackage{graphicx}% Include figure files
\usepackage{dcolumn}% Align table columns on decimal point
\usepackage{bm}% bold math
\usepackage{epsfig}
\usepackage{psfrag}
\usepackage{subfigure}

\begin{document}

\title{Trainspotting: Extraction and Analysis of Traffic and Topologies of Transportation Networks}

\author{Maciej Kurant, Patrick Thiran}

\date{\today}

\begin{abstract}
The knowledge of real-life traffic pattern is crucial for good understanding and analysis of
transportation systems. This data is quite rare. In this paper we propose an algorithm for
extracting both the real physical topology and the network of traffic flows from timetables of
public mass transportation systems. We apply this algorithm to timetables of three large
transportation networks. This enables us to make a systematic comparison between three different
approaches to construct a graph representation of a transportation network; the resulting graphs
are fundamentally different. We also find that the real-life traffic pattern is very heterogenous,
both in space and traffic flow intensities.
\end{abstract}


 \maketitle

\section{Introduction}\label{sec:Intro}
In the recent years, studies of transportation networks have drawn a substantial amount of
attention in the physics community. The graphs derived from the physical infrastructure of such
networks were analyzed on the examples of a power grid~\cite{Watts98,AlbertUSAPowerGrid}, a railway
network~\cite{Indian03,Gastner04b}, road
networks~\cite{Latora04dual,Gastner04,Rosvall05,Latora05primal,Cardillo05}, pipeline
network~\cite{Gastner04b} or urban mass transportation systems
\cite{Latora01,Latora02,Stations04,Sienkiewicz05,Vragovic05}. These studies have one important
feature in common - they focus exclusively on the topology of the network, and they do not take
into account the real-life traffic pattern. This makes the view very incomplete, because carrying
traffic is the ultimate goal of every transportation system. Facing the lack of real-life traffic
data, some authors try to estimate the traffic pattern based exclusively on the topology. Probably
the most common load estimator is \emph{betweenness} (used e.g.,
in~\cite{Goh01,Szabo02,Bollobas04,Holme02,Zhao04,Motter04}), which assumes that each pair of nodes
exchanges the same amount of traffic. But the real-life traffic patterns are in fact very
heterogenous, both in space and traffic flow intensities. Therefore the most important nodes and
edges from a topological point of view might not necessarily carry the most traffic.
In~\cite{KurantLayeredNetworks} we show that in typical transportation networks the correlation
between the real load and the betweenness is very low.
%Moreover, the real traffic pattern seems very difficult to be predicted by
%any topology based metric.
Therefore it is essential for some applications to know the real traffic pattern.\\
Interestingly, the networks of traffic flows were studied separately, see the example of flows of
people within a city~\cite{Chowell03}, and commuting traffic flows between different
cities~\cite{Montis05}. These studies, in turn, neglect the underlying physical topology, making
the analysis incomplete. For instance, it is impossible to detect the most loaded physical edges,
which might have a crucial meaning for the resilience of the system. A comprehensive view of the
system often requires to analyze both layers (physical and traffic) together.

Unfortunately, the data sets including both physical topology and traffic flows are rather sparse,
and difficult to get. In this paper we propose an approach to extract the physical structure and
the network of traffic flows from \emph{timetables}. Timetables of trains, buses, trams, metros and
other means of mass transportation (henceforth called \emph{vehicles}) are publicly available. They
provide us with the available connections and their times. Timetables also contain the information
about the physical structure of the network and the traffic flows in it, but, as we show later,
they often require a nontrivial preprocessing to be revealed.

\begin{figure*}
    \psfrag{a}[][]{{\footnotesize A}}
    \psfrag{b}[][]{{\footnotesize B}}
    \psfrag{c}[][]{{\footnotesize C}}
    \psfrag{d}[][]{{\footnotesize D}}
    \psfrag{e}[][]{{\footnotesize E}}
    \psfrag{f}[][]{{\footnotesize F}}
    \psfrag{g}[][]{{\footnotesize G}}
    \psfrag{h}[][]{{\footnotesize H}}
    \psfrag{s}[l][]{shortcut}
    \psfrag{x}[l][]{{\footnotesize Line 1}}
    \psfrag{y}[l][]{{\footnotesize Line 2 (express)}}
    \psfrag{z}[l][]{{\footnotesize Line 3}}
\centering
    \subfigure[${}\;$Three train lines]{\includegraphics[width=0.21\textwidth]{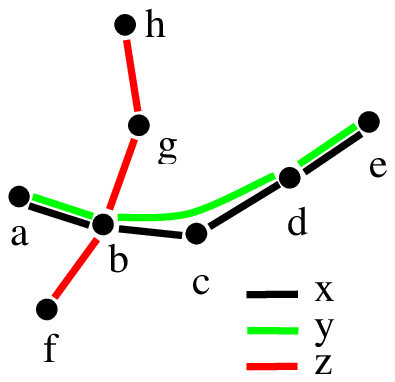}}
    \hspace{1cm}
    \subfigure[${}\;$space\---of\---changes]{\includegraphics[width=0.18\textwidth]{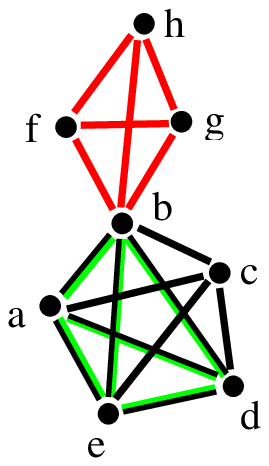}}
    \hspace{1cm}
    \subfigure[${}\;$space\---of\---stops]{\includegraphics[width=0.21\textwidth]{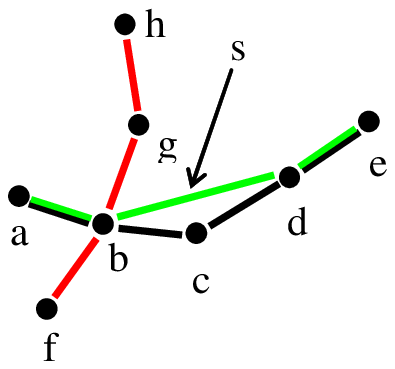}}
    \hspace{0.7cm}
    \subfigure[${}\;$space\---of\---stations]{\includegraphics[width=0.21\textwidth]{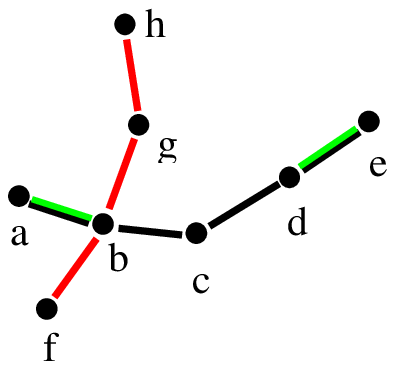}}
    \caption{(Color online) An illustration of the transportation network topology in three spaces.
    (a)~The routes of three vehicles. The route of Line~2
    passes through node~C on the way from B to D, but the vehicle does not stop there.
    (b)~The topology in space\---of\---changes. Each route results in a clique.
    An edge is indicated by two colors, when it originates from two routes, but is merged into a single link.
    (c)~The topology in space\---of\---stops. The ``shortcut'' B-D is a legitimate edge in this space.
    (d)~The topology in space\---of\---stations. This graph reflects the topology of the real-life infrastructure.}
\label{fig:Spaces}
\end{figure*}

\section{Spaces and the difficulty of the problem}
In order to position our contribution in the range of works in the field, we begin with a
systematic definition of the topology of transportation systems. The set of nodes is defined by the
set of all stations (train stations, bus stops, etc). It is not obvious, however, what should be
interpreted as an edge. Its choice depends on what we want to be reflected by the topology of the
physical graph. In the literature there are essentially three approaches that define three
different `spaces': here we call them `space\---of\---changes', `space\---of\---stops' and
`space\---of\---stations':\\
% \footnote{Two of these spaces were described and compared
%in~\cite{Sienkiewicz05} under the names \emph{space~L} and \emph{space~P}. To obtain a complete and
%systematic picture, here we introduce a
%third one}:\\
In \textbf{space\---of\---changes}, two stations are considered to be connected by a link when
there is at least one vehicle that stops at both stations. In other words, all stations used by a
single vehicle are fully interconnected and form a clique. This approach neglects the physical
distance between the stations. Instead, in the resulting topology, the length of a shortest path
between two arbitrary stations $A$ and $B$ is the \emph{number of changes} of mean of
transportation one needs to get from $A$ to $B$~\footnote{In this sense, a graph in
space\---of\---changes is closely related to the \emph{dual} interpretation of urban road
networks~\cite{Latora04dual,Kalapala05,Rosvall05}, where streets (of a given name) map to nodes,
and intersections between streets map to links between the nodes. In a transportation network in
space\---of\---changes, the length of a shortest path is the number of changes of mean of
transportation, whereas the length of a shortest path in a dual graph of a city is the number of
changes of streets on the way from the starting point to destination.}. This approach was used
in~\cite{Indian03,Stations04,Sienkiewicz05}; in the latter the authors used the
term \emph{space~P}.\\
In \textbf{space\---of\---stops}, two stations are connected if they are two consecutive stops on a
route of at least one vehicle~\cite{Sienkiewicz05}. Here the length of a shortest path between two
stations is the minimal \emph{number of stops} one needs to make. Note that the number of stations
traversed on the way might be larger, because the vehicles do not necessary stop on all of them.\\
In \textbf{space\---of\---stations}, two stations are connected only if they are physically
directly connected (with no station in between). This reflects the topology of the real-life
infrastructure. Here, the length of a shortest path between two stations is the minimal
\emph{number of stations} one has to traverse (stopping or not). This approach was used
in~\cite{Latora01,Latora02,Gastner04b,Vragovic05}.

In Fig.~\ref{fig:Spaces} we give an illustration of the three spaces. It is easy to see that the
graph in space\---of\---stations is a subgraph of the graph in space\---of\---stops, which in turn
is a subgraph of the graph in space\---of\---changes.

The topologies in space\---of\---changes and space\---of\---stops can be directly obtained from
timetables. In space\---of\---changes, for each vehicle, we fully connect all stations it stops at.
Then we simplify the resulting graph by deleting multi-edges. In space\---of\---stops, we connect
every two consecutive stops in routes of vehicles. As shown in Fig.~\ref{fig:Spaces}c, the topology
in space\---of\---stops can have shortcut links that do not exist in the real-life infrastructure.
These shortcuts should be eliminated in the space\---of\---stations topology, which makes it more
challenging to obtain. To the best of our knowledge, the only work on extracting the real physical
structure (the topology in space\---of\---stations) from timetables was done in the context of
railway networks in the PhD dissertation of Annegret Lebers~\cite{PhD_Trains}. The proposed
solution first obtains the physical graph in space\---of\---stops. Next, specific structures in the
initial physical graph, called \emph{edge bundles}, are detected. The Hamilton
paths\footnote{\emph{Hamilton path} is a path that passes through every vertex of a graph exactly
once} within these bundles should indicate the real (non-shortcut) edges. Unfortunately, the bundle
recognition problem turned out to be NP-complete. The heuristics proposed in~\cite{PhD_Trains}
result in a correct real/shortcut classification of 80\% of edges in the studied graphs. The
approach we propose in this paper is based on simple observations that were omitted
in~\cite{PhD_Trains}. This results in a much simpler and more effective algorithm.

%We explain this on the following example illustrated in left-hand side of
%Fig.~\ref{fig:PathReplace}. The train~1 stops successively at the stations
%$A\!\!-\!\!B\!\!-C\!\!-\!\!D$. In a straightforward but naive approach we interpret every direct
%connection as a physical link (rail track); this results in three physical links: $A\!\!-\!\!B$,
%$B\!\!-\!\!C$ and $C\!\!-\!\!D$. The train~2, an express one, stops at $A\!\!-\!\!B\!\!-\!\!D$. The
%naive approach would add another physical link, $B\!\!-\!\!D$, to the previous set. Nevertheless,
%it is very likely that the second train passes through the station $B$, but does not stop there. So
%the link $B\!\!-\!\!D$ is a \emph{shortcut link} that does not actually exist in the physical
%topology, which means that the naive approach is not sufficient.

%However, in a general and more frequent case, where the routes of vehicles are constrained to
%physical infrastructure (roads, rail tracks, power lines), we need some nontrivial processing of
%timetables in order to be extract the physical graph and the traffic pattern.

\section{Related work}
Timetables have been used as a data source for a network construction
in~\cite{Indian03,Sienkiewicz05}. However, the topologies obtained in these works were either in
space\---of\---changes or in space\---of\---stops; neither of them reflected the real-life
infrastructure. Moreover, the real traffic patterns were not considered in these studies. This is
understandable, because it is difficult to interpret a traffic flow in spaces~of~changes and~stops.
Does the ``traffic'' on a shortcut link have any physical meaning? We know that this traffic
actually traverses other non-shortcut links that exist in reality. In contrast, in
space\---of\---stations, the traffic flows have clear, unambiguous and natural interpretation.

%timetables USED FOR SURE IN Gastner04,

Another class of networks that can be constructed with the help of timetables are airport
networks~\cite{Barrat04,Hufnagel04,Gastner04,Guimera05}. There, the nodes are the airports, and
edges are the flight connections. The weight of an edge reflects the traffic on this connection,
which can be approximated by the number of flights that use it during one week. In this case, both
the topology and the traffic information are \emph{explicitly} given by timetables. This is because
the routes of planes are not constrained to any physical infrastructure, as opposed to roads for
cars or rail-tracks for trains. So there are no ``real'' links and ``shortcut'' links. In a sense
all links are real, and the topologies in space\---of\---stops and in space\---of\---stations
actually coincide.

Inferring the space\---of\---stations topology from timetables becomes simple also in another
special case, where the vehicles stop at each station they traverse (e.g., in many subway
networks). This naturally eliminates the shortcuts, making the topologies in space\---of\---stops
and~stations identical. This is not true in a general case, with both local and express vehicles.

In the reminder of this paper, we introduce necessary notation in Section~\ref{sec:Notation}. Next,
in Section~\ref{sec:Algotithm} we give an algorithm that extracts the real physical structure (a
topology in space\---of\---stations) and the network of traffic flows from timetables. In
Section~\ref{sec:Comparison} we test our algorithm on timetables of three large transportation
networks at three different scales: city, country and continent. We also analyze the resulting
physical topologies and compare them with those obtained by alternative approaches.
%; we discover fundamental differences between them.
Finally, in Section~\ref{sec:Conclusion} we conclude the paper.

%We evaluate the accuracy of the algorithm on real and artificially generated datasets.

%EXPAND TO SHOW THE DIFFICULTY OF THE PROBLEM

%Discuss THE SIZES OF THESE NETWORKS.!!!

%DISCUSS THE TRAFFIC FLOWS

\section{Notation}\label{sec:Notation}
%\begin{figure*}[!htb]
%    \psfrag{o}[r][]{Train 1 (local):\hspace{0.6cm}}
%    \psfrag{t}[r][]{Train 2 (express):\hspace{0.6cm}}
%    \psfrag{p}[r][]{Resulting physical graph:\hspace{0.6cm}}
%    \psfrag{r}[c][]{{\large Original}}
%    \psfrag{m}[c][]{{\large Modified}}
%    \psfrag{s}[l][]{shortcut}
%    \psfrag{a}[][]{{\footnotesize A}}
%    \psfrag{b}[][]{{\footnotesize B}}
%    \psfrag{c}[][]{{\footnotesize C}}
%    \psfrag{d}[][]{{\footnotesize D}}
%\includegraphics[width=0.7\textwidth]{figures/PathReplace.eps}
%\caption{An example run of the algorithm, with two trains. The first train is local, stops at all
%stations. The second train is express, and passes through station C without stopping. Based on the
%two routes, we infer the physical topology. The naive approach (on the left) results in a physical
%topology in space\---of\---stops, i.e., with a shortcut B-D that does not exist in reality. In order to
%eliminate this undesirable link we detect it in the route of the train~2 and replace with the
%corresponding path B-C-D from the train~1 (right-hand side of the figure). Now, the resulting
%physical graph is in space\---of\---stations, i.e., has no shortcuts.} \label{fig:PathReplace}
%\end{figure*}

\subsection{Two layers}
We follow the \emph{two-layer framework} introduced in~\cite{KurantLayeredNetworks}. The
lower-layer topology is called a \emph{physical graph} $G^\phi=(V^\phi, E^\phi)$, and the
upper-layer topology is called a \emph{logical graph} $G^\lambda=(V^\lambda, E^\lambda)$. We assume
that the sets of nodes at both layers are identical, i.e., $V^\phi\equiv V^\lambda$, but as a
general rule, we keep the indexes $\phi$ and $\lambda$ to make the description unambiguous. Let
$N\!\!=\!\!|V^\phi|\!\!=\!\!|V^\lambda|$ be the number of nodes. Every logical edge
$e^\lambda=\{u^\lambda,v^\lambda\}$ is mapped on the physical graph as a path $M(e^\lambda)\subset
G^\phi$ connecting the nodes~$u^\phi$ and~$v^\phi$, corresponding to $u^\lambda$ and $v^\lambda$.
(A path is defined by the sequence of nodes it traverses.) The set of paths corresponding to all
logical edges is called a \emph{mapping} $M(E^\lambda)$ of the logical topology on the physical
topology.

In the field of transportation networks the undirected, unweighted physical graph $G^\phi$ captures
the topology of the physical infrastructure (i.e., in space\---of\---stations), and the weighted
logical graph $G^\lambda$ reflects the undirected traffic flows. Every logical edge $e^\lambda$ is
created by connecting the first and the last node of the corresponding traffic flow, and by
assigning a weight $w(e^\lambda)$ that represents the intensity of this flow. The mapping
$M(e^\lambda)$ of the edge $e^\lambda$ is the path taken by this flow.

%ADD FIGURE??

%WHAT IF WE HAVE TWO VEHICLES CONNECTING THE SAME END-STATIONS, BUT BY DIFFERENT ROUTES??

\subsection{Timetable data}
We take a list of all vehicles departing in the system within some period (e.g., one weekday).
Denote by $R=\{r_i\}_{i=1..|R|}$ the list of routes followed by these vehicles, where $|R|$ is the
total number of vehicles. A route $r_i$ of $i$th vehicle is defined by the list of nodes it
traverses. Note that since there are usually more vehicles (than one) following the same path on
one day, some of the routes may be identical.

\section{Algorithm}\label{sec:Algotithm}

The algorithm has three phases. In the first one, initialization, based on the set of routes $R$,
we create the set of nodes $V^\phi=V^\lambda$ and the physical topology $G^\phi_{stop}=(V^\phi,
E^\phi_{stop})$ in space\---of\---stops. In the second, main phase, the sets $R$ and
$E^\phi_{stop}$ are iteratively refined by detecting and erasing the shortcut links in the physical
graph $G^\phi_{stop}$, resulting in the physical topology $G^\phi_{stat}=(V^\phi, E^\phi_{stat})$
in space\---of\---stations. Finally, in the third phase, we group the vehicles with identical
routes, and obtain the logical graph $G^\lambda$ and the mapping~$M(E^\lambda)$ of the logical
edges on the physical graph~$G^\phi_{stat}$. We describe below each phase separately.

\subsection{Phase 1 - initialization}
%$V^\phi=V^\lambda=\bigcup_i V(r_i)$
In this phase we interpret every two consecutive nodes in any route $r_i\in R$ as directly
connected. Consequently, we connect these nodes with a link, which can be written as
$$E^\phi_{stop}=\bigcup_{i=1..|R|}E(r_i)$$
where $E(r_i)\;$ is the set of all pairs of adjacent nodes in $r_i$ (i.e., all edges in $r_i$).
This results in the physical topology $G^\phi_{stop}=(V^\phi, E^\phi_{stop})$ in
space\---of\---stops.

\subsection{Phase 2 - deleting shortcuts}
In this phase, at each iteration, we detect a shortcut in the set of physical edges, delete it, and
update all routes $r_i$ that use this shortcut. Denote by $e^\phi_{(1)}, e^\phi_{(2)}$ the two
end-nodes of $e^\phi$, and by Rev$(P_{e^\phi})$ the reversed version of $P_{e^\phi}$ (the sequence
from the last node to the first one). The algorithm is as follows:

\begin{enumerate}
    \item $E^\phi_{stat}:=E^\phi_{stop}$

    \item Find a tuple $(e^\phi, r_i)$ such that $e^\phi$ is a shortcut for $r_i$:\\
     $e^\phi_{(1)}\in r_i$ \quad and\quad $e^\phi_{(2)}\in r_i$\quad and\quad $e^\phi\notin E(r_i)$.

    \item IF no $(e^\phi,r_i)$ found THEN RETURN $E^\phi_{stat}$ and $R$.

    \item $P_{e^\phi}$ := subpath of $r_i$ from $e^\phi_{(1)}$ to $e^\phi_{(2)}$

    \item FOR all $r_j\in R$ DO:\\
%     $\bullet$ Replace in $r_i$ every occurrence of $(e^\phi(1), e^\phi(2))$ with $P_{e^\phi}$\\
%     $\bullet$ Replace in $r_i$ every occurrence of $(e^\phi(2), e^\phi(1))$ with Reverse$(P_{e^\phi})$
     $\bullet$ If $(e^\phi_{(1)}, e^\phi_{(2)})\in r_j$ THEN replace it with $P_{e^\phi}$\\
     $\bullet$ If $(e^\phi_{(2)}, e^\phi_{(1)})\in r_j$ THEN replace it with Rev$(P_{e^\phi})$

    \item $E^\phi_{stat}:=E^\phi_{stat}\setminus\{e^\phi\}$

    \item GOTO~2
\end{enumerate}

In Step~2, we look for a physical link that is a shortcut. We declare a physical link~$e^\phi$ to
be a shortcut, if there exists a route $r_i\in R$, such that $e^\phi$ connects two
\emph{non}consecutive nodes in $r_i$. For example, in Fig.~\ref{fig:Spaces}c, $e^\phi=\{B,D\}$ is a
shortcut because it connects two not neighboring nodes in the route $r_1$ of Line~1. If no physical
edge can be declared a shortcut, the algorithm quits in Step~3, returning~$E^\phi_{stat}$ and~$R$.
Otherwise, in Step~4, we find the path $P_{e^\phi}$ that this shortcut should take. In
Fig.~\ref{fig:Spaces}c this path is $P_{e^\phi}=(B,C,D)$. In Step~5, we update the set of routes
$R$ by replacing every shortcut link $e^\phi$ in every route using it with the corresponding path
$P_{e^\phi}$. In our example, the updated route of Line~2 becomes $r_2=(A,B,C,D,E)$. It is thus
identical to the route of Line~1. Finally, in Step~6 we delete the shortcut~$e^\phi$ from the
physical graph. We iterate these steps until no shortcut is found (Step~2). The resulting physical
graph~$G^\phi_{stat}=(V^\phi, E^\phi_{stat})\subset G^\phi_{stop}$, is a graph in
space\---of\---stations.

\subsection{Phase 3 - grouping the same routes together}
Finally, based on the list $R$ of routes updated in phase~2, we find groups of vehicles that follow
the same path (in any direction). Each such group defines one edge $e^\lambda$ in the logical
graph; $e^\lambda$ connects the first and the last node of the route. The number of vehicles that
follow this route becomes the weight $w(e^\lambda)$ of the logical edge~$e^\lambda$; the route
itself becomes the mapping $M(e^\lambda)$ of $e^\lambda$ on the physical graph.\\ Denote by
$r_{i(first)}, r_{i(last)}$ the first and the last nodes in $r_i$, and by $E(M(e^\lambda))$ the set
of all physical edges in the mapping of $e^\lambda$. Now, Phase~3 can be stated as follows:
\begin{enumerate}
    \item $E^\lambda=\emptyset$, $M=\emptyset$

    \item FOR $i=1$ TO $|R|$ DO:\\
        $\bullet$ $e_i^\lambda=\{r_{i(first)}, r_{i(last)}\}$\\
        $\bullet$ IF $e_i^\lambda\in E^\lambda$ THEN $w(e_i^\lambda) := w(e_i^\lambda)+1$\\
        ${}\;\,$ ELSE $E^\lambda\!=\!E^\lambda\bigcup \{e^\lambda_i\}$,\quad $M(e^\lambda_i)\!=\!r_i$,\quad $w(e^\lambda_i)\!=\!1$

    \item $E^\phi_{stat}=\bigcup_{e^\lambda\in E^\lambda}E(M(e^\lambda))$
\end{enumerate}
In the example in Fig.~\ref{fig:Spaces}, after phase~2 the routes of Line~1 and Line~2 become
identical; therefore in phase~3 they are grouped together defining a logical edge
$e^\lambda_1=\{A,E\}$ with the weight $w(e^\lambda_1)=2$ and the mapping
$M(e^\lambda_1)=(A,B,C,D,E)$. A second logical edge is $e^\lambda_2=\{F,H\}$ with
$w(e^\lambda_2)=1$ and $M(e^\lambda_2)=(F,B,G,H)$.

\subsection{Accuracy of the algorithm}
There are potential sources of mistakes and inaccuracies in our approach. First, the links that we
delete as being shortcuts, might actually exist in reality. However, a comparison of the results of
our algorithm with the real maps (see Section~\ref{sec:Comparison}) reveals very few
differences, which means that this source of failures occurs very rarely in real data sets.\\
A second problem lies in the estimation of the traffic pattern. Interpreting the routes of trains,
buses, trams, metros, etc, as traffic flows gives us a picture at a low level of granularity. We
view every vehicle as a traffic unit, regardless of its size or the number of people it carries.
Moreover, people usually use these vehicles only on a portion of its total journey, not from the
first to the last station. Clearly, the vehicle routes are the result of an optimization process
that take into account many factors, such as people's demand, continuity of the path, traveling
times and availability of stock. However, we believe that they reflect well the general direction
and intensity of travels, and we take a vehicle as a basic traffic unit. After all, these are the
vehicles that appear on the roads and cause traffic, not the people they transport.

\begin{figure*}
\centering
    \subfigure[${}\;$Physical graph $G^\phi_{change}$ in space\---of\---changes]{\includegraphics[width=0.45\textwidth]{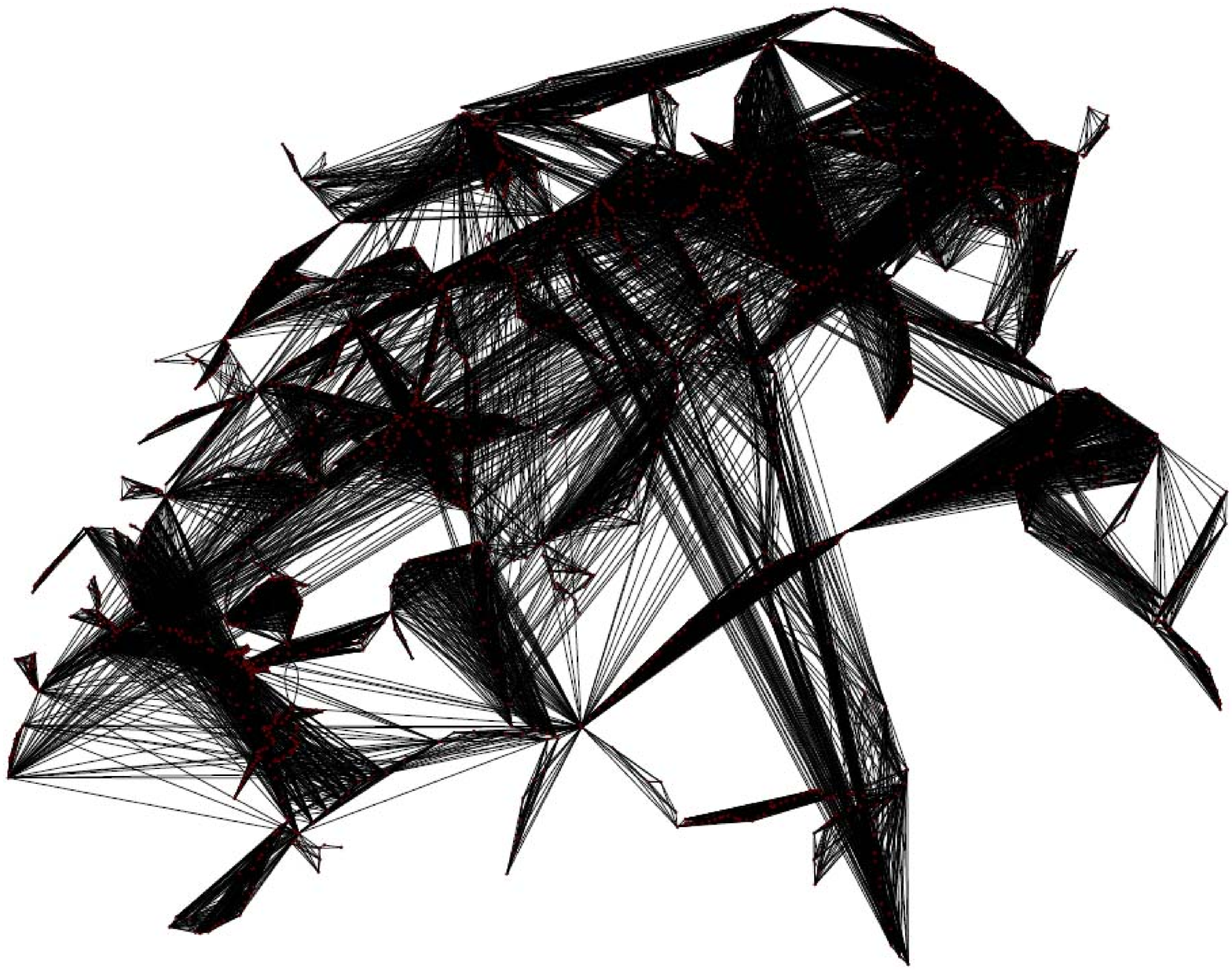}}
    \subfigure[${}\;$Physical graph $G^\phi_{stop}$ in space\---of\---stops]{\includegraphics[width=0.45\textwidth]{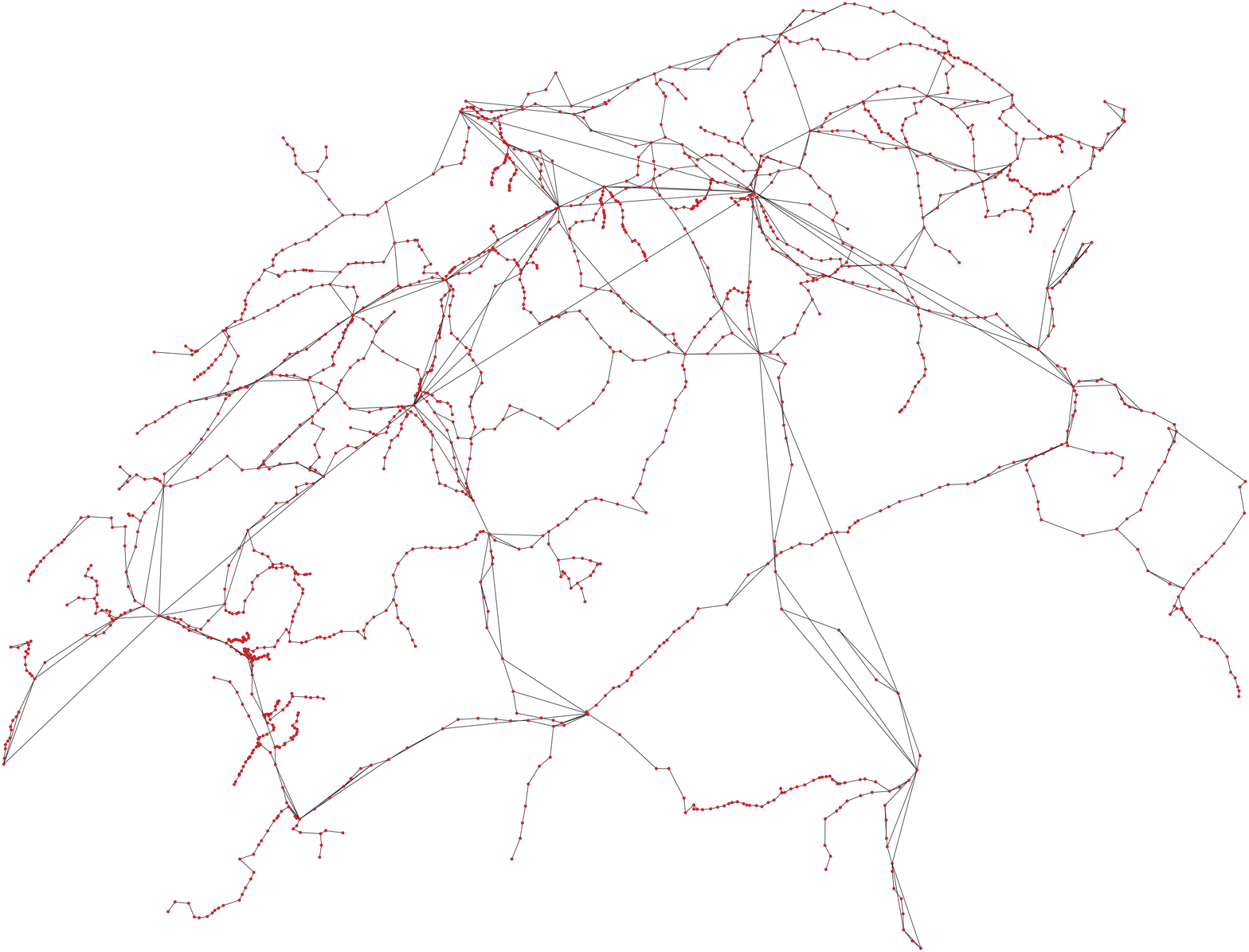}}\\
    \subfigure[${}\;$Physical graph $G^\phi_{stat}$ in space\---of\---stations]{\includegraphics[width=0.45\textwidth]{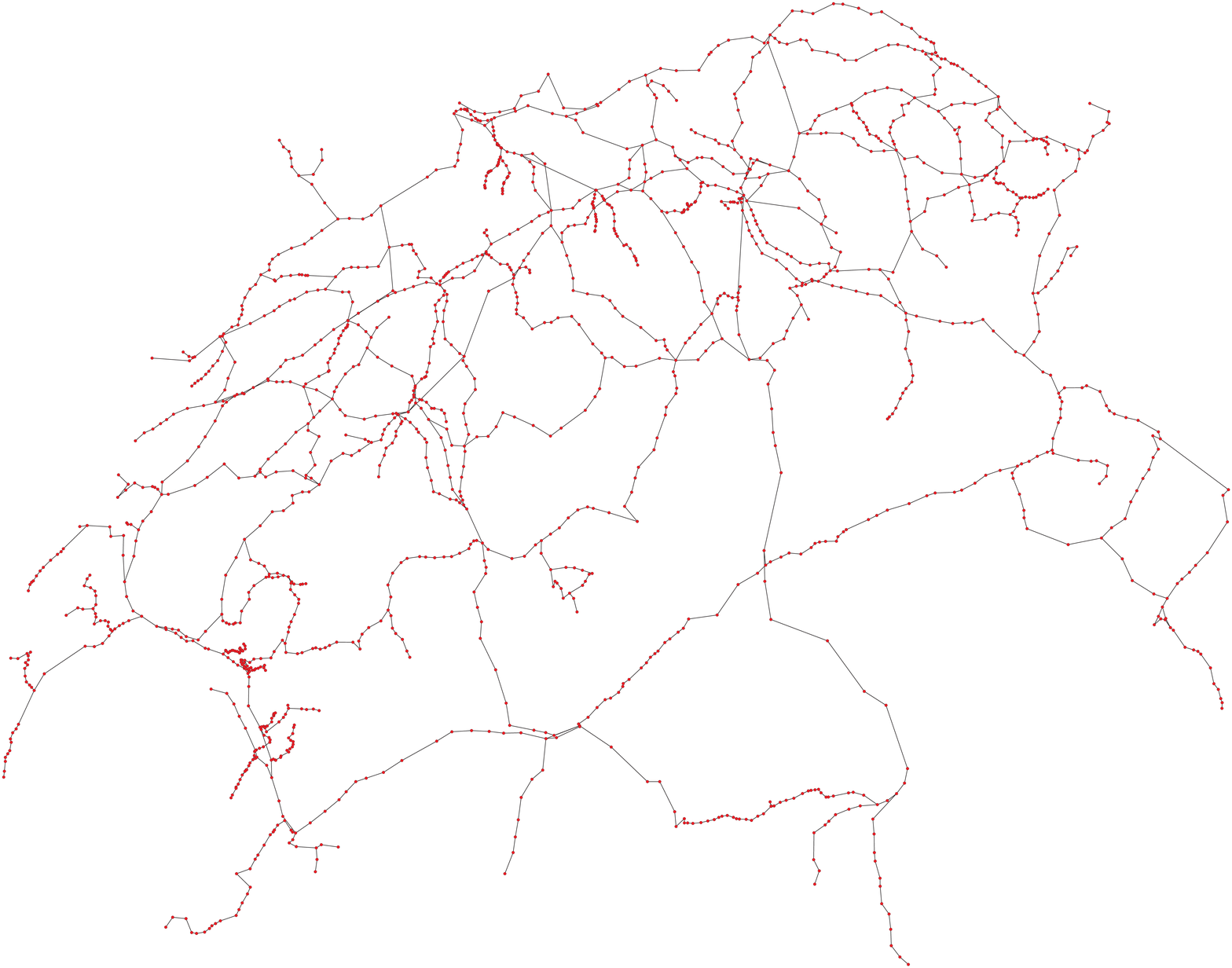}}
    \subfigure[${}\;$Real physical map]{\includegraphics[width=0.45\textwidth]{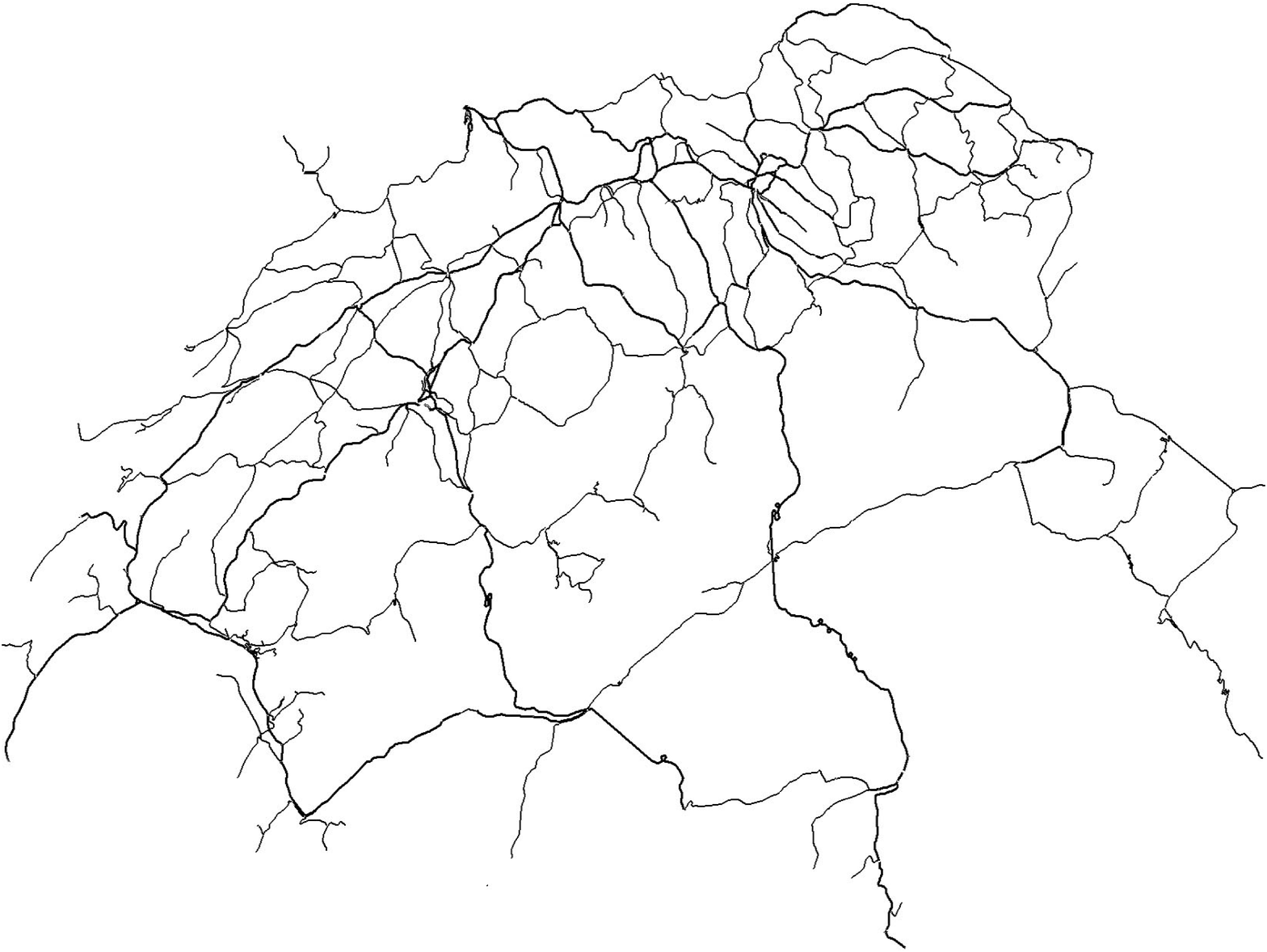}}\\
    \subfigure[${}\;$Logical graph]{\includegraphics[width=0.5\textwidth]{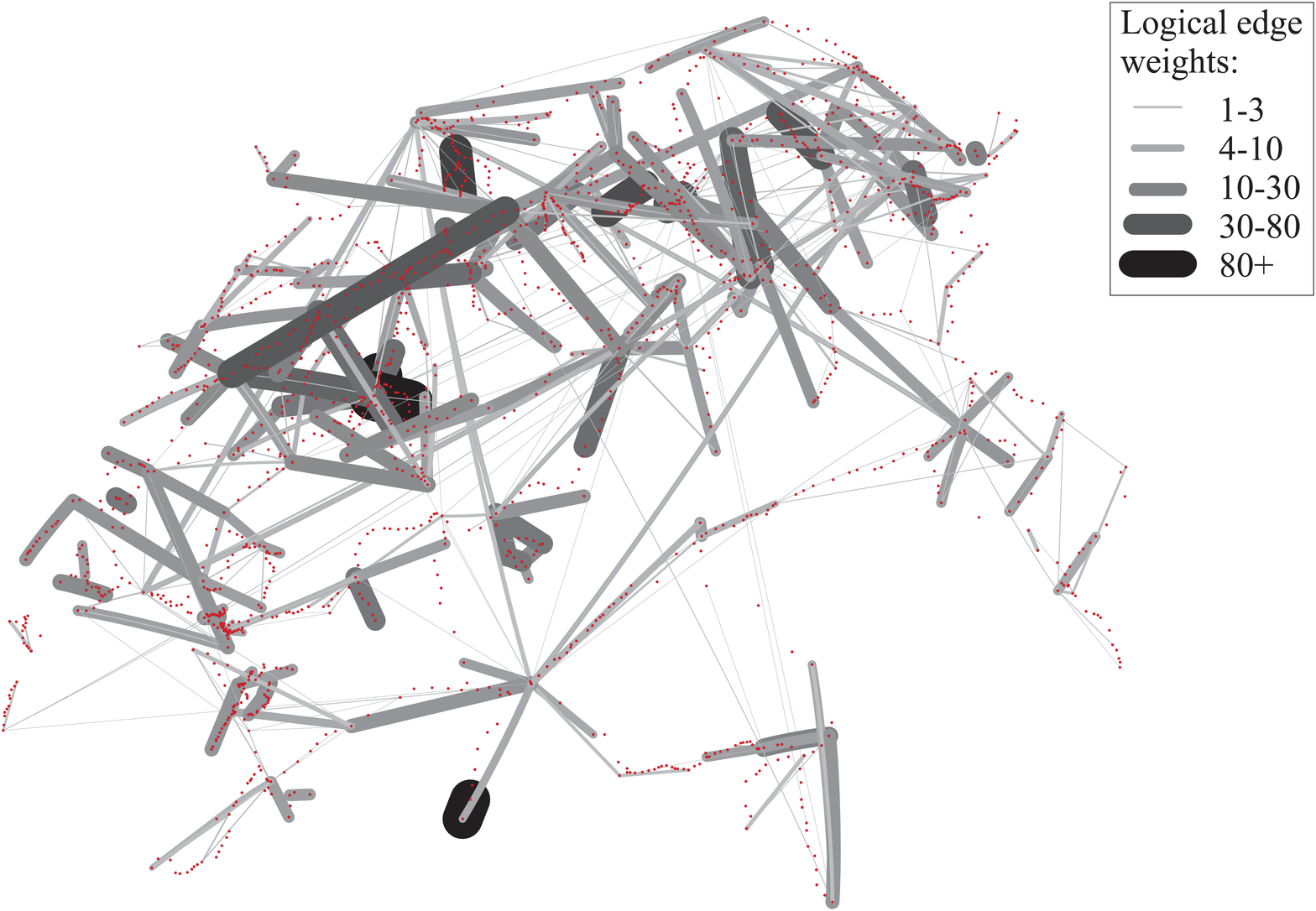}}
    \caption{The railway network in Switzerland (CH). (a,b,c) Physical graphs in space\---of\---changes, stops and stations,
    respectively. (d) The real map of the rail tracks in Switzerland. (e) The logical graph.
    Every edge connects the first and the last station of a particular train route; its
    weight reflects the number of trains following this route in any direction.} \label{fig:GraphComparison}
\end{figure*}

\begin{table*}[!htb]
\begin{tabular}{|l|r|r|r|r|c|r|r|r|r|r|}

  \hline
  \multicolumn{3}{|c|}{General} & \multicolumn{2}{c|}{Traffic} & \multicolumn{6}{c|}{Physical graph}\\
  \hline
  \multicolumn{1}{|c|}{Dataset} &Area $[km^2]$&\multicolumn{1}{c|}{$N$}& \multicolumn{1}{c|}{$|R|$}&\multicolumn{1}{c|}{$|E^\lambda|$}&
  \multicolumn{1}{c|}{\hspace{0.1cm}Space} & \multicolumn{1}{c|}{$|E^\phi|$} & \multicolumn{1}{c|}{$\langle k^\phi\rangle$}&
  \multicolumn{1}{c|}{$d^\phi$} & \multicolumn{1}{c|}{$\langle l^\phi\rangle$}& \multicolumn{1}{c|}{$c^\phi$}\\
  \hline
  && &&&changes  &\hspace{0.1cm}78437&\hspace{0.1cm} 102.3 &4& 2.3&0.6829 \\%& 0.0668\\
  WA (Warsaw)&480&1533&25'995&221&stops & 2249& 2.9 &76&19.0& 0.1681\\%&0.0019\\
  && &&&stations& 1832 &2.4&  90 & 28.1 & 0.0092\\%&0.0016\\
  \hline
  && &&&changes &19827&24.6&8&3.6&0.9095\\%&0.0153\\
  CH (Switzerland)&41'300&1613&6'957&505&stops& 1922&2.4&61&16.3& 0.0949\\%&0.0015\\
  && &&&stations & 1680 & 2.1&136 & 46.6 & 0.0004 \\%& 0.0013\\
  \hline
  &&&&&changes  & 88329 & 36.4& 8 & 3.7  & 0.7347 \\%&0.0075\\
  EU (Europe)&2'081'000&\hspace{0.1cm}4853&\hspace{0.1cm}60'775&\hspace{0.1cm}6703&stops &8600 &3.5& 48 & 12.6 & 0.3401 \\%&0.0007\\
  && &&&stations &5765 &2.4& \hspace{0.25cm}184 & \hspace{0.3cm}50.9 &\hspace{0.3cm}0.0129 \\%&0.0005\\
  \hline
\end{tabular}
\caption{The studied datasets. ``Area'' is the surface occupied by the region covered by the
network. $N$ is the number of nodes (stations/stops). $|R|$ is the total number of vehicles
departing in the network during one weekday. $|E^\lambda|$ is the number of edges in the logical
graph (number traffic flows); it is much smaller than $|R|$, because the vehicles following the
same route are grouped together in phase~3 of our algorithm. All the remaining parameters are
computed for the physical graphs $G^\phi$: $|E^\phi|$ is the number of edges, $\langle
k^\phi\rangle$~is the average node degree, $d^\phi$ stands for the diameter, $\langle
l^\phi\rangle$ is the average shortest path length, and $c^\phi$ is the clustering coefficient. }
\label{tab:Comparison}
\end{table*}

\section{A study of three real-life networks}\label{sec:Comparison}

In this section we apply our algorithm to extract the data from the timetables of three examples of
transportation networks, with sizes ranging from city to continent. As an example of a city, we
take the mass transportation system (buses, trams and metros) of Warsaw~(WA), Poland; its
timetables are available at~\cite{ZKM_Warszawa}. At a country level, we study the railway network
of Switzerland~(CH). Finally, we investigate the railway network formed by major trains and
stations in most countries of central Europe~(EU)\footnote{In the EU data set, Paris has originally
several stations that are not directly connected between each other. Following the approach
in~\cite{Gastner04b}, we merged them into one common node.}. The timetables of both CH and EU
networks are available at~\cite{CFF}. The basic parameters of the data sets and of the resulting
graphs can be found in Table~\ref{tab:Comparison}.

% The sizes of the studied networks vary from $N=1613$ to $4853$.

This section is organized as follows. First, we focus on a particular data set in order to study
the performance of our algorithm. Next, we analyze and compare the physical graphs originating from
all three data sets in each of the considered spaces. Finally, we focus our attention on the
logical graphs and traffic flows extracted by our algorithm.

\subsection{An example: The railway network of Switzerland (CH)}
As an illustration, let us consider more closely the railway network of Switzerland~(CH).
%We have collected the routes
%of trains of the following types: CIS, E, EC, EN, IC, ICE, ICN, IR, R, RE, S, TGV.
According to our timetable, on a typical weekday there are $|R|=6957$ different trains that follow
$|E^\lambda|=505$ different routes (usually there is more than one train following the same route
during one day). Our data contains $N=1613$ stations in Switzerland, together with their physical
coordinates. In Fig.~\ref{fig:GraphComparison} we present the graphs obtained from this data set.
The physical graphs in the three spaces are shown in Figs.~\ref{fig:GraphComparison}abc. The graph
in space\---of\---stations  was obtained with the help of the algorithm introduced in the previous
section.
%; it needed three iterations to converge.
The number of vertices is the same in all three spaces. The number of edges in
space\---of\---changes, $|E^\phi_{change}|=19827$, is much larger than in the other two spaces.
Although at first sight the physical graphs in space\---of\---stations and in space\---of\---stops
look comparable, the latter has a number of (nonexisting in reality) shortcut links. For a visual
verification of correctness of our algorithm, we show in Fig.~\ref{fig:GraphComparison}d the real
map of the Swiss railway system; we observe only minor differences between~(c) and~(d). Finally, in
Fig.~\ref{fig:GraphComparison}e, we present the logical graph that reflects the traffic flows in
the network.
%The edge weights are equal to the number of trains following a given route, which is
%extracted from the timetables.
This graph is very heterogenous both in the weights of edges and in
the layout of traffic.

\subsection{The physical graph in three spaces}
%All physical graphs are connected.
How does the choice of space affect the topology? We study in this section the physical graphs in
the three spaces with respect to the basic metrics often used in the analysis of complex networks.

\subsubsection{Diameter $d^\phi$, and average shortest path length $\langle l^\phi\rangle$}
The average shortest path length $\langle l\rangle$ is computed over the lengths of shortest paths
between all pairs of vertices. The diameter $d$ is the longest of all shortest path lengths. These
parameters are usually closely related.

The diameters and average shortest path lengths of the graphs in space\---of\---stations are large,
and scale roughly as $\sqrt{N}$ with the number of nodes~$N$. This is typical of many planar,
lattice-like infrastructure networks embedded in a two dimensional space.

The graphs in space\---of\---stops have about $10-15\%$ more edges than their counterparts in
space\---of\---stations. The difference is not large, and one could possibly expect similar values
of the diameter and the average shortest path length. However, these $10-15\%$ edges are
fundamentally different from typical edges in space\---of\---stations; they are shortcut links. It
was shown in~\cite{SmallWorldsByWatts} that the diameter of a graph is very sensitive to the
existence of shortcuts. Even a relatively small number of shortcuts can dramatically bring down the
diameter and the average shortest path length. We observe this phenomenon in our graphs. For
instance, in the EU data set, the diameter drops about four times, from $d^\phi=184$ in
space\---of\---stations to 48 in space\---of\---stops. Similarly, the average shortest path length
drops by roughly the same factor. Therefore, the shortcut edges, although not very numerous,
play a very important role and make the graphs in space\---of\---stops very different from those in space\---of\---stations.\\
This effect is not so strongly pronounced in the WA data set. The underlying reason is the
relatively short length of shortcuts (usually 2 hops), which was shown to affect the diameter only
to a small extent~\cite{Kleinberg00}.

Finally, the graphs in space\---of\---changes have very small diameters and average shortest path
lengths. This is mainly because of their high density (number of edges).
%However, even these dense
%graphs retain some structure - they have clustering coefficients much higher than a corresponding
%random graph and the diameter still larger then them [some works on the relation between clustering
%coefficient and diameter???]

\subsubsection{Node degree $k$}
\begin{figure}
  \psfrag{x1}[c][]{$k^\phi$}
  \psfrag{y1}[c][]{$Pr(k^\phi)$}
  \psfrag{WA}[c][]{WA}
  \psfrag{CH}[c][]{CH}
  \psfrag{EU}[c][]{EU}
\centering
    \subfigure[${}\;$space\---of\---changes ($G^\phi_{change}$)]{\includegraphics[width=0.23\textwidth]{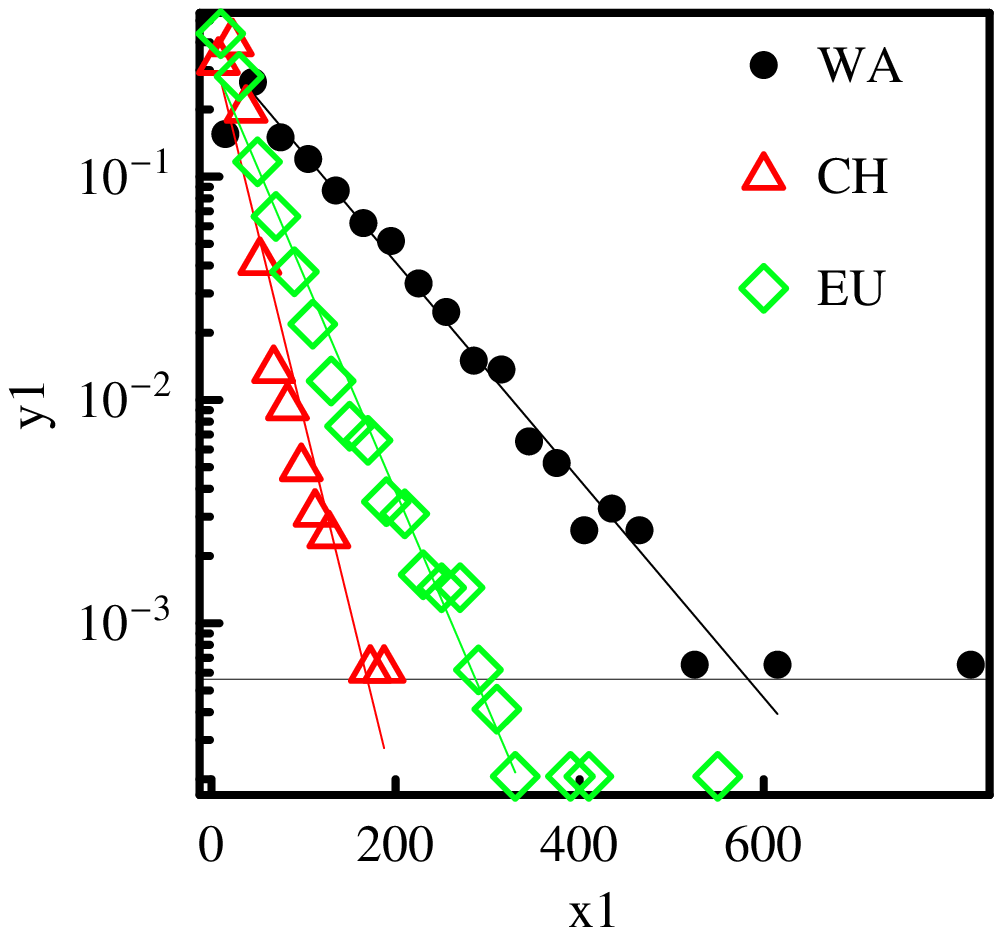}}\hspace{0.3cm}
    \subfigure[${}\;$space\---of\---stops ($G^\phi_{stop}$)]{\includegraphics[width=0.23\textwidth]{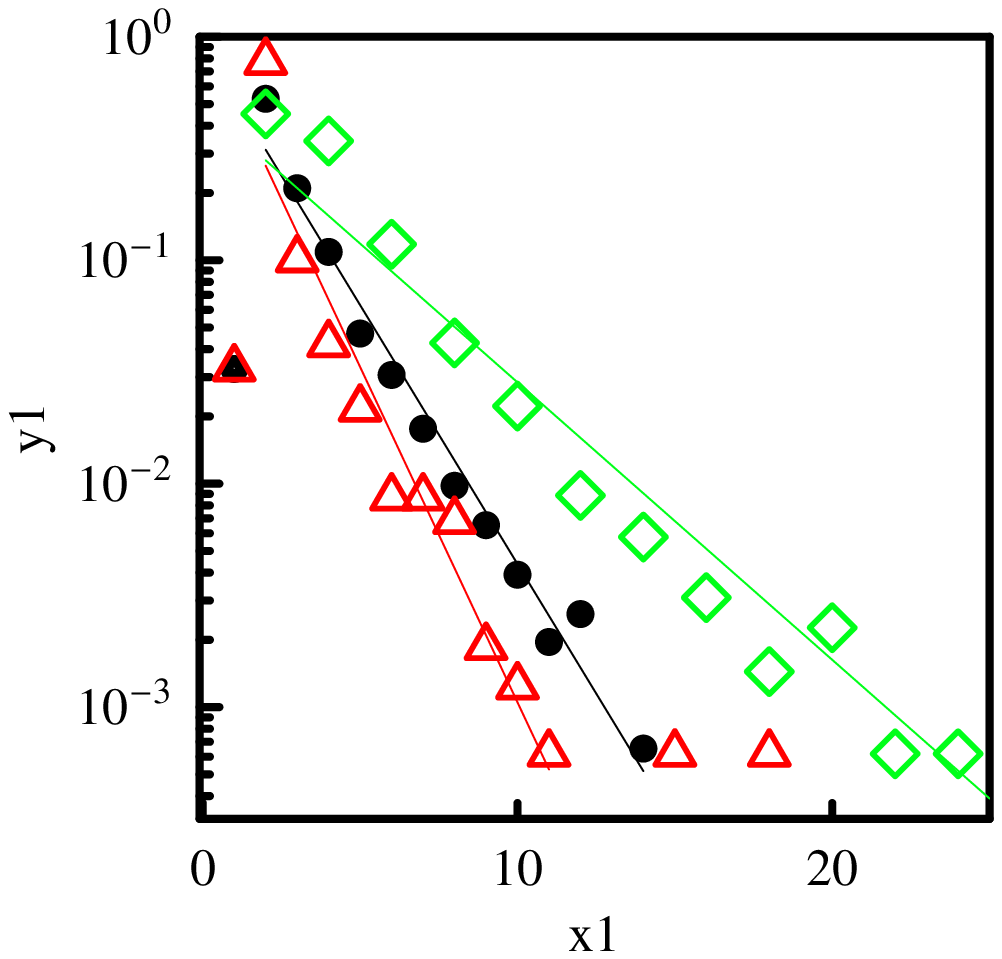}}\\
    \subfigure[${}\;$space\---of\---stations ($G^\phi_{stat}$)]{\includegraphics[width=0.23\textwidth]{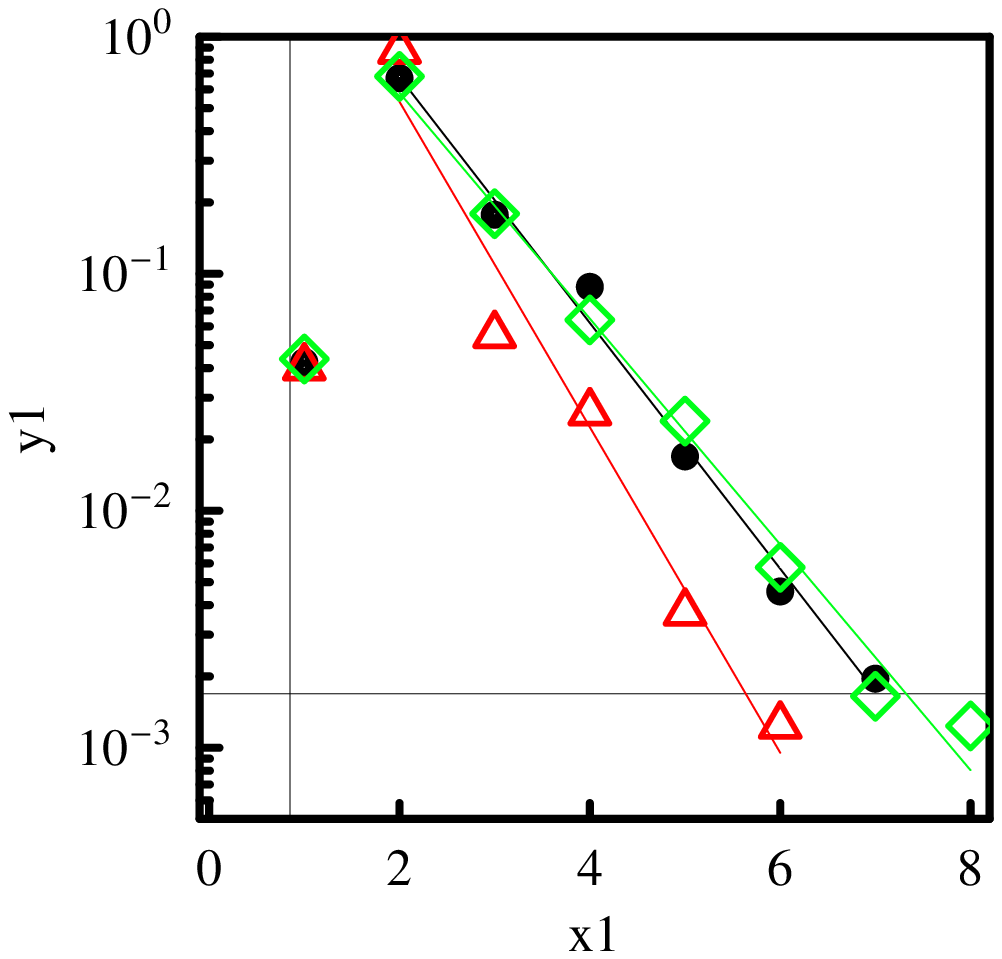}}\hspace{0.3cm}
    \subfigure[${}\;$space\---of\---stops, log-log scale]{\includegraphics[width=0.23\textwidth]{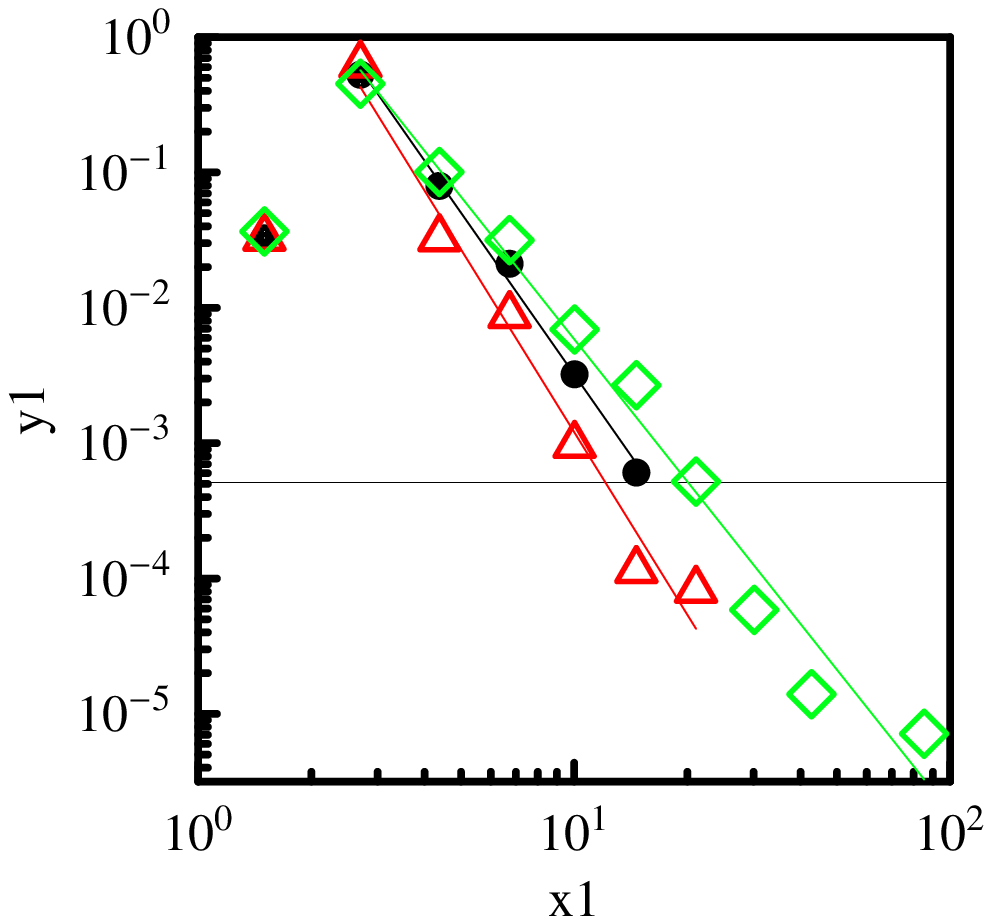}}
    \caption{Node degree distributions in physical graphs in the three spaces, for the data sets WA, CH and EU.
    Plots (a-c) use a semi-logarithmic scale, plot (d) uses a log-log scale. If necessary, the data
    is lin-binned or log-binned, accordingly.} \label{fig:DegreeComparison}
\end{figure}

The node degree distributions in all three spaces are plotted in a semi-logarithmic scale in
Fig.~\ref{fig:DegreeComparison}abc. Additionally, for space\---of\---stops, we plot the degree
distributions in a log-log scale (Fig.~\ref{fig:DegreeComparison}d), because it is not obvious
which fit is better, exponential or power law (it was also pointed out in~\cite{Sienkiewicz05}).
For the other two spaces we observe a clear linear trend indicating the exponential behavior. This
was expected in space\---of\---stations, because the degree distribution of many infrastructure
networks was shown to be narrow (here one decade) and to decay exponentially (see e.g., power lines
in~\cite{Strogatz01}). In space\---of\---stations the vast majority of nodes have degree equal to
two, indicating long segments of stations without junctions.

%Note that in space\---of\---changes, by construction, the degree distribution is strongly related to the
%lengths of original (!!!!) vehicle routes. We elaborate on that in Section~\ref{subsec:traffic}.

%(\begin{minipage}[c]{0.2cm} \includegraphics[width=1\textwidth]{red.eps}\end{minipage})

\subsubsection{Clustering coefficients $c$}
We have studied the clustering coefficients $c$ defined as a probability that two randomly chosen
neighbors of a node are also direct neighbors of each other~\cite{SmallWorldsByWatts}.

The clustering coefficient of topologies in space\---of\---changes are very high, which is a direct
consequence of a very high density and existence of many cliques. What is more interesting is that
in all three data sets, the clustering coefficient in space\---of\---stops is 1-2 orders of
magnitude larger than in space\---of\---stations. As in the case of the graph diameter, here again
the shortcut links turn out play a
very important role in the topology.%to greatly affect the metric of clustering.

%Much higher than of random graph in P , L , (K? - maybe smaller than of random?)

%filtered out

\subsection{Traffic flows and the logical graph}\label{subsec:traffic}
\begin{figure}
    \subfigure[${}\;$WA]{\includegraphics[width=0.155\textwidth]{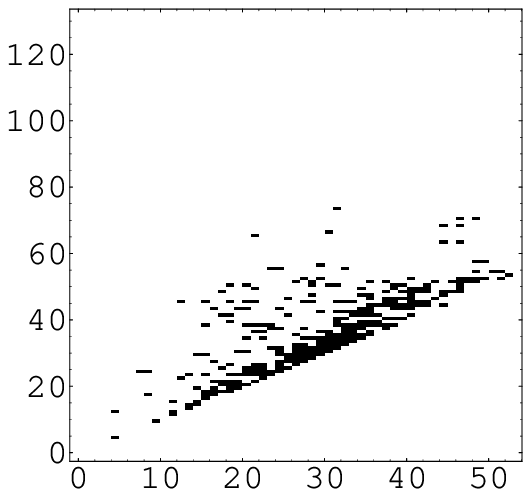}}
    \subfigure[${}\;$CH]{\includegraphics[width=0.155\textwidth]{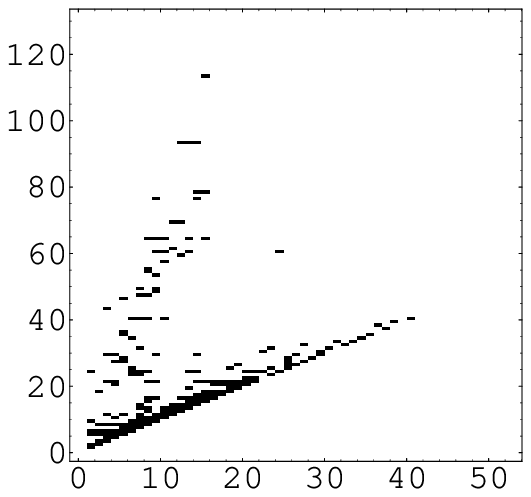}}
    \subfigure[${}\;$EU]{\includegraphics[width=0.155\textwidth]{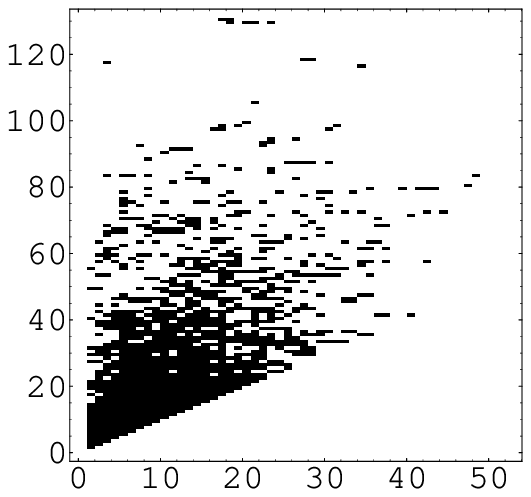}}
    \caption{The lengths of original timetable routes ($x$~axis) versus these lengths after
    the application of our algorithm ($y$~axis). All three data sets are drawn in the same scale.
    } \label{fig:densities}
\end{figure}
Now we turn our attention to the traffic that flows in our networks. We extracted this scarce data
with the help of the algorithm introduced in this paper. As we argued before, the interpretation of
traffic flowing through networks in space\---of\---changes and~stops is rather cumbersome.
Therefore we restrict our analysis to the traffic flows traversing the physical graph in
space\---of\---stations.

In Fig.~\ref{fig:densities} we compare the lengths of traffic flows before and after application of
our algorithm. A new traffic flow can be either equal in length to the original one (if no shortcut
was detected on its path), or longer. We observe that for all three data sets, there is a
significant number of flows that become longer. In some cases this increase in length is by as much
as $10$~times. Generally, the longer the original flow is, the less extended it gets during a run
of our algorithm. This is expected, because a long flow in a timetable usually corresponds to a
local train that stops at all stations (i.e., uses no shortcuts).

\begin{figure}
    \psfrag{x1}[c][]{$k^\lambda$}
    \psfrag{y1}[c][]{$Pr(k^\lambda)$}
    \psfrag{x2}[c][]{$s^\lambda$}
    \psfrag{y2}[c][]{$Pr(s^\lambda)$}
    \psfrag{x3}[c][]{$w(e^\lambda)$}
    \psfrag{y3}[c][]{$Pr(w(e^\lambda))$}
    \subfigure[${}\;$Node
    degrees]{\includegraphics[width=0.23\textwidth]{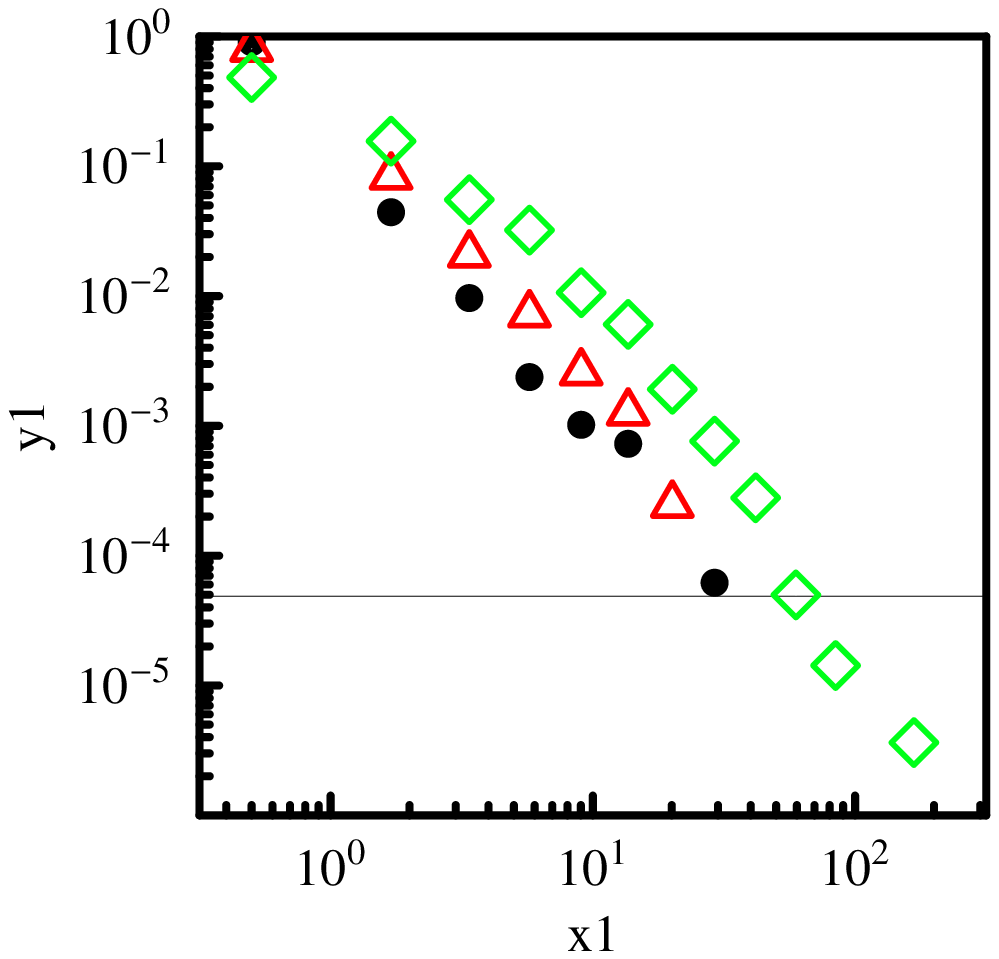}}\hspace{0.3cm}
    \subfigure[${}\;$Node strengths
    ]{\includegraphics[width=0.23\textwidth]{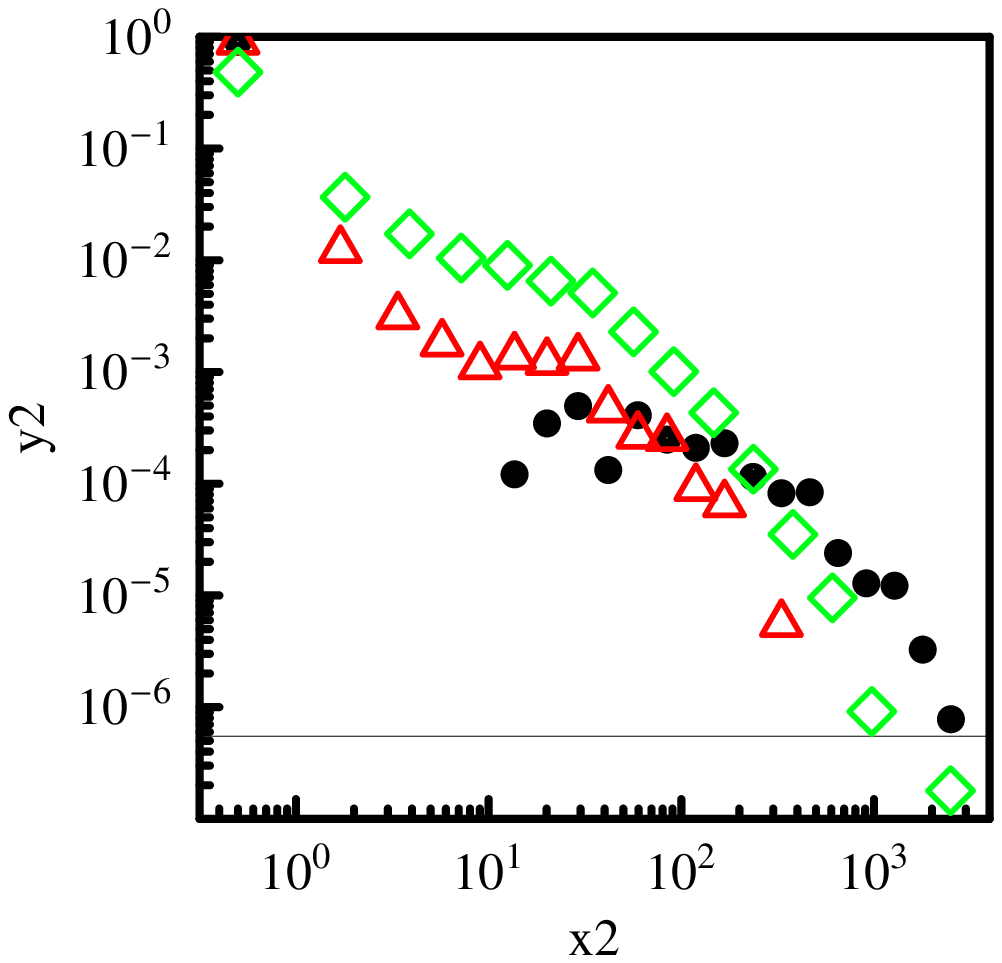}}\\
    \subfigure[${}\;$Edge weights]{\includegraphics[width=0.23\textwidth]{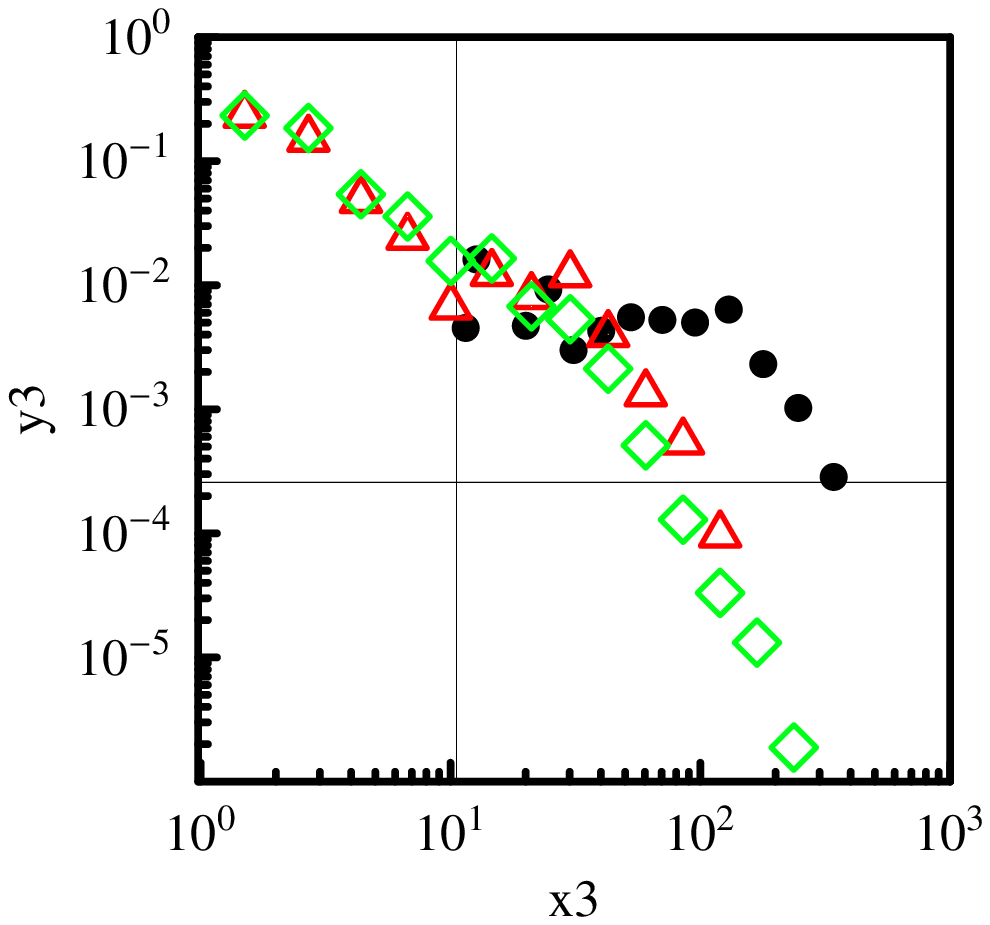}}
    \caption{Properties of logical graphs.
    (a)~Node degree distribution. Many nodes are isolated - they represent intermediate stations on
    which no train starts or terminates its journey. The isolated nodes we represent here as having ``degree'' equal
    to~$0.1$.\hspace{0.2cm}
    (b)~Node strength distribution. \hspace{0.2cm}
    (c)~Edge weight (traffic flow intensities) distribution.\hspace{0.2cm}
    All data are log-binned and plotted in a log-log scale.} \label{fig:logGraph}
\end{figure}

In Fig.~\ref{fig:logGraph} we present basic distributions measured for logical graphs in the three
data sets. Recall that the edges in a logical graph reflect the traffic flows. Therefore, the node
degree~$k^\lambda$ is the number of \emph{different} connections starting/ending at the
corresponding station (Fig.~\ref{fig:logGraph}a). The strength~$s^\lambda$ of a node is the sum of
the weights of neighboring edges~\cite{Barrat04}; here it is the number of \emph{all} connections
starting/ending at this station (Fig.~\ref{fig:logGraph}b). Finally, the weight~$w(e^\lambda)$  of
a logical edge is the traffic flow intensity (Fig.~\ref{fig:logGraph}c).\\
All three distributions are
%wide (cover two or more decades) and
heavily right-skewed meaning that
there is a small number of nodes/edges with very high values of the observed parameter. We conclude
that the real-life traffic patterns are very heterogenous, both in space (node degree and strength)
and traffic flow intensities. This was shown in~\cite{KurantLayeredNetworks} to be the reason of
high unpredictability of load distribution in transportation networks.

%Note that the values of weights can be very diverse: from 1 to several hundreds. The high values
%are especially visible for the WA data set, because it is based mainly on city busses, which are
%often much more frequent than trains.

%flow length distribution (weighted and unweighted. Unweighted should be highly correlated with
%degree distr in space\---of\---changes)
%
%some references to PRL paper.

\section{Conclusions}\label{sec:Conclusion}

The knowledge of real-life traffic pattern is crucial in the analysis of transportation systems.
This data is usually much more difficult to get than the pure topology of a network. In this paper
we have proposed an algorithm for extracting both the physical topology and the network of traffic
flows from timetables of public mass transportation systems.
%This gives a comprehensive view of the studied systems.
We have applied our algorithm to three large transportation networks. This enabled us to make a
systematic comparison between three different approaches (or ``spaces'') to construct a graph
representation of a transportation network. The resulting physical topologies are very different.
In particular, the seemingly similar graphs in space\---of\---stops and in space\---of\---stations,
turn out to be very different in terms of basic graph-theory metrics such as diameter, average
shortest path length, clustering coefficient and node degree distribution. This is due to the
existence of shortcut links in space\---of\---stops. Our algorithm detects and eliminates these
shortcuts, and extracts the topology in space\---of\---stations. Only this graph reflects the
real-life physical infrastructure that is used by the traffic flows, gets congested or can be prone
to failures or susceptible to attacks. In contrast, the edges in space\---of\---changes and in
space\---of\---stops are somewhat ``virtual,'' and the notion of traffic in these graphs is
unclear, if at all makes any sense. What is important, the results are consistent across three
different scales of the studied networks (city, country, continent).

This work has several possible directions for the future. For instance, the knowledge of real
traffic pattern allows us to revisit the error and attack tolerance~\cite{Albert00} of
transportation systems, which might look completely different when focussing on traffic instead of
on topology. Another direction would be to exploit additional information available in some
timetables. For instance, in our data sets CH and EU, we also know the geographical coordinates of
the nodes. They fall therefore in the category of \emph{spatial networks} that have been recently
intensively studied~\cite{Gastner04b,Gastner04,Masuda05,Petermann05,Barrat05,Cardillo05}. In
particular, we think that incorporating the real traffic pattern in the models can help
understanding the processes that govern the evolution of spatial networks.

Finally, we note that the data will be soon available at~\cite{KurantWWW}.

%The analysis presented in this paper was basic. Many more interesting results can be extracted from
%this data set, especially if we take the two layers into account.

The work presented in this paper was financially supported by grant DICS~1830 of the Hasler
Foundation, Bern, Switzerland.

%WA timetables were taken from the same source as in~\cite{Sienkiewicz05}, so we can really compare
%to their paper.

\bibliographystyle{unsrt}
%%\bibliography{../literatureMK}

\end{document}